# Toward Neuroadaptive Communication Interfaces: A Framework for Personalized Speech and Language Support

*Carlos Bruno*[1], *Luis Ricardo Santos*[1,2]

[1] Faculty of Psychology and Education Sciences, University of Coimbra
[2] Amazon AGI

## Abstract

Speech and language technologies have expanded access to communication, but the support they provide is often fixed. The same linguistic or visual aid may help one user and hinder another, or become less useful as task demands change. This project proposes a framework for adapting language presentation to individual abilities and ongoing interaction. It separates access to a message from the estimation of user needs and from the selection of support, including sentence segmentation, pictograms, and optional spoken output. Behavioral feedback provides the primary evidence for adaptation; electroencephalography is considered an additional, uncertainty-aware source of information rather than a direct measure of comprehension. We specify a staged evaluation comparing fixed, behavior-adaptive, and neuroadaptive interfaces. The central question is whether neural information improves comprehension and communication efficiency beyond a strong behavioral baseline.

**Index Terms**: neuroadaptive interfaces, assistive communication, multimodal language support

## 1. Introduction

Communication technologies have made substantial progress in overcoming limitations of conventional speech and motor output. Intracranial speech brain-computer interfaces (BCIs) have demonstrated rapid text decoding and multimodal speech-related output in individual participants with severe paralysis [1], [2]. At the same time, clinical translation still raises questions of robustness, training, long-term use, and the division of learning between users and decoders [3]. Non-invasive speech decoding research addresses a different point on the feasibility spectrum as many studies tried decoding different aspects of speech [4], [5], [6], [7]. Such demonstrations are important steps toward neural access to communication, but they should not be confused with ready-to-use, open-vocabulary communication systems.

Access, however, is only one part of a successful interaction. Once a message is available, a person still needs to understand, formulate, or select it. The relevant difficulties vary widely across acquired motor impairments, language disorders, and developmental conditions; these are not interchangeable clinical populations. Even within a single population, identical presentations may place different demands on perception, language, and working memory, as is the case for attention deficit hyperactivity disorder (ADHD), dyslexia, and autism per example [8], [9], [10]. This does not establish an neural marker of comprehension, but it illustrates why a diagnosis alone is an incomplete specification of an individual's language-processing profile.

Adaptive presentation offers one possible response. In a small supervised study involving learners with special educational needs, sentence segmentation and pictorial scaffolds did not produce a uniform pattern of benefit; richer displays could also introduce coordination costs [11]. Meanwhile, automated mapping of textual concepts to pictograms has shown the technical feasibility of generating such supports across several languages, although language-specific repository coverage and semantic ambiguity remain relevant [12]. Thus, the ability to generate support and the decision to use it are separate problems.

Neuroadaptive systems extend adaptive interaction by using brain activity to inform the interface, typically through passive estimates of an aspect of the user's state [13], [14]. Recent EEG-informed conversational systems have demonstrated adaptive text generation, but improvements in engagement did not necessarily translate into improved learning outcomes [15]. This motivates a more circumscribed question in communication support: when, if ever, does neural information improve the choice of a linguistic aid beyond what can be inferred from behavior?

We outline a project to answer this question. Its contribution is not a new speech decoder or a claim of clinical efficacy. It is a modular distinction between communication access, support selection, and the evidence used for adaptation, accompanied by an evaluation designed to isolate the added value of neural measurements.

## 2. Framework and Design Principles

The proposed interface is organized around a message, a user model, and a constrained adaptation policy (Fig. 1). The message can arrive through ordinary text input, speech recognition, or an alternative access method. Neural speech decoding is a possible longer-term input route, not a prerequisite for the initial experiments. Crucially, the neural activity used to decode an intended message and the neural activity used to estimate cognitive state serve different functions, require different validation, and need not be measured with the same system.

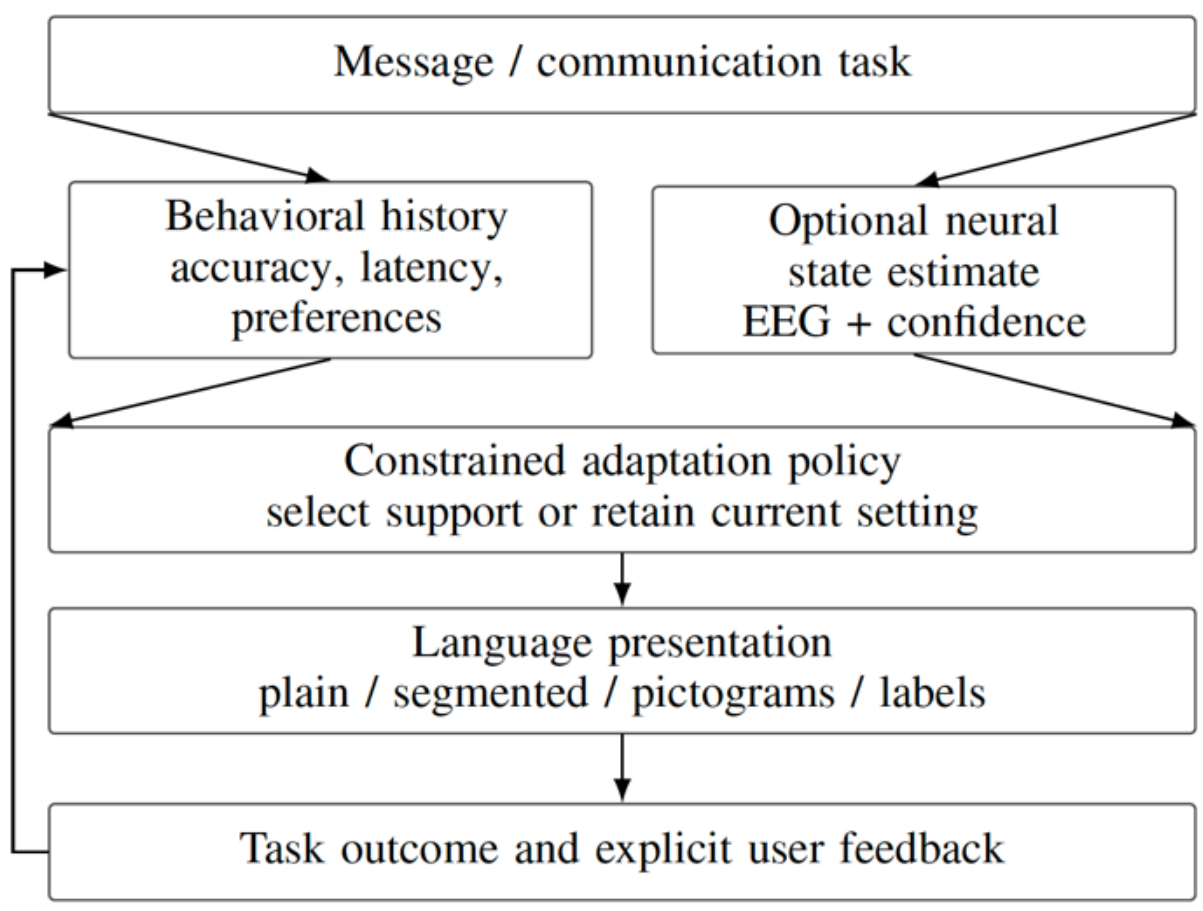

Figure 1: *Proposed separation of message access, state estimation, and selection of language support. The neural branch is optional and is tested against the behavioral branch alone. The figure describes a planned system, not an implemented clinical device.*

For the first prototype, a message is presented in one of four controlled formats: unmodified text; sentence-segmented text; segmented text with selected pictograms; or the latter with keyword labels. These conditions are deliberately close to previously studied scaffolds [11], allowing the work to concentrate on selection rather than on inventing new forms of assistance. Optional read-aloud output can be added in a later version. The underlying wording is held constant between the four initial formats, so that changes in performance are not attributed to simultaneous, uncontrolled text rewriting.

### 2.1. User model and adaptation policy

The user model has two timescales. A slowly updated profile summarizes language assessments, preferred formats, and accessibility settings. A task-level record contains correctness, response time, requests for help, and recent selections. Neural information, if collected, enters as a calibrated estimate with an explicit measure of uncertainty. It does not determine whether a response has been understood.

Let $a_t$ denote the presentation format selected for the next item and $h_t$ the available behavioral history. A behavior-adaptive policy chooses $a_t$ from $h_t$ and the user's stated constraints. Its neuroadaptive counterpart additionally uses a neural estimate $\hat{z}_t$ only if signal quality and calibration criteria are satisfied. Both policies have the same set of actions and are trained under the same information budget. A low-confidence neural estimate triggers the behavioral policy rather than an enforced interface change. To avoid disruptive fluctuations, adaptation initially occurs between messages, not during the reading of a sentence. Users and clinicians can override any selection.

We favor an interpretable, regularized policy over unconstrained reinforcement learning at this stage. A small feasibility sample will not support complex individual policies, and the system must remain auditable: its choice should be traceable to recent performance, preferences, or a properly validated neural estimate.

## 3. Research Plan

### 3.1. Phase 1: Content and interface validation

An initial prototype will implement the four presentation formats and log the timing and source of every adaptation. French and Spanish are suitable first languages for a controlled pilot; extension to other languages will depend on content coverage and qualified review. Automatically selected pictograms will be checked for semantic adequacy, polysemy, and unintended changes in meaning. Stimuli will be short functional messages and brief passages with comprehension questions, not arbitrary generated text. A small group of speech-language and accessibility professionals will review materials before participant testing.

### 3.2. Phase 2: Behavioral feasibility study

We envisage a pilot of approximately 30 adults covering a range of reading and language abilities, including a subgroup reporting developmental reading difficulties. The number is a feasibility target, not a power calculation or a claim that data have been collected. Clinical diagnoses, if used for subgroup analyses, will be verified rather than inferred from performance. Eligibility and accessible testing procedures will be finalized with clinical collaborators and ethics approval.

Participants will encounter equivalent, counterbalanced sets of messages under a fixed-format interface and a behavior-adaptive interface. A short calibration block will estimate baseline performance and preferences. The adaptive rule will then be locked for evaluation on held-out messages, avoiding an overly optimistic assessment on the same material used to tune it. Different message sets, rather than repetitions of the same questions, will be used across conditions. The initial objective is to estimate feasibility, within-person variability, adherence, and effect-size uncertainty; a larger, separately powered study would be needed for efficacy claims.

### 3.3. Phase 3: Does EEG add information

In a subsequent EEG-equipped subset, the same tasks will compare the fixed condition, the behavior-adaptive policy, and the identical policy supplemented by EEG-derived information. A separate calibration task will be used to test whether the chosen EEG features capture a reproducible state relevant to the task. Spectral measures may be considered, but no single frequency band will be presumed to measure comprehension. Artifacts, eye movements, and changes in visual layout must be addressed because they can correlate with both the selected scaffold and the neural features.

Incremental value will be tested on held-out sessions. A neural-label permutation control will help identify apparent gains caused by chance or model complexity; signal-quality failures and missing data will be reported rather than selectively excluded. The comparison is meaningful only if neural measurements change decisions and those changes improve user outcomes. Accurate decoding of a surrogate EEG label, by itself, will not meet that criterion.

## 4. Evaluation and Interpretation

The primary behavioral outcome will be comprehension accuracy, accompanied by response time to characterize the speed–accuracy trade-off. Prespecified secondary measures will include subjective effort, requests for assistance, adaptation frequency, overrides, interface latency, and semantic errors in pictogram selection. Communication rate may become relevant for a later message-composition or BCI-access task, but should not substitute for comprehension in the proposed first study.

Analyses will compare policies within participant while accounting for repeated items; mixed-effects models and effect estimates with uncertainty intervals are appropriate starting points. Any group-by-policy interaction in the pilot will be treated as exploratory. For the neural condition, a positive result would require improvement over the behavior-adaptive comparator, not merely over a static display. Conversely, no added neural benefit would be informative: it would indicate that the cost and complexity of EEG are not justified for this task under the tested conditions.

Several boundaries are important. Neither a language-processing difference observed in dyslexia nor a neural workload estimate can automatically be generalized to people living with locked-in syndrome. The initial prototype addresses support selection for language comprehension; integrating an implantable speech decoder would constitute a separate clinical development and validation effort. Adaptation also creates risks of excessive assistance, unstable display changes, and reduced user agency. We therefore constrain permissible changes, preserve manual control, and treat language-specific content safety and data privacy as prerequisites, not secondary refinements.

## 5. Conclusion

The project proposes a testable route from generic assistive communication toward personalization informed by individual behavior and, where justified, neural state. Its main premise is deliberately conditional: neural measurements should be added only when they provide reliable information that improves the user's actual communication outcome. This separation of access, adaptation, and evaluation makes it possible to build on progress in speech BCIs and multilingual language support without assuming that their integration is already clinically effective.